\documentclass{cimento}

\usepackage{graphicx}
\usepackage{amsmath}
\usepackage{amssymb}
\title{New physics searches in flavour physics}
\author{David M. Straub \from{ins:sns}}
\instlist{\inst{ins:sns} Scuola Normale Superiore and INFN, Piazza dei Cavalieri 7, 56126 Pisa, Italy}
\PACSes{%
\PACSit{12.60.-i}{Models beyond the standard model}
\PACSit{13.20.He}{Decays of bottom mesons}
}
\begin{document}

\maketitle

\begin{abstract}
The origin of flavour and CP violation is among the most important open questions in particle physics. Imminent results from the LHC as well as planned dedicated flavour physics experiments might help to shed light on this puzzle. This talk concentrates on the NP sensitivity of the rare $B$ decays $B_{s,d}\to\mu^+\mu^-$ and $B\to K^*\mu^+\mu^-$ and of the CP violating phase in $B_s$ mixing. A brief summary of a supersymmetric model with interesting signatures in the flavour sector is presented.
\end{abstract}

\section{Introduction}

\noindent
The tremendous progress in experimental flavour physics facilitated by the $B$ factories, the Tevatron and other experiments shows that the CKM picture of flavour and CP violation describes the data well, both for tree and loop-induced processes.
Since the Standard Model (SM) is only an effective theory valid up to some energy scale $\Lambda$, which is yet to be determined but is expected to lie in the TeV region from naturalness considerations, this success is a challenge for theories of new physics (NP). Assuming generic flavour and CP violating couplings, the bounds on $\Lambda$ from $\Delta F=2$ processes exceed 10\,000 TeV \cite{Isidori:2010kg}.
On the other hand, given that a theoretical understanding of the hierarchical structure of quark masses and CKM mixings is still lacking, this paradox might simply indicate that the flavour structure of NP is related to the SM one. If in fact the Yukawa couplings, already present in the SM, are the {\em only} sources of breaking of the flavour symmetry, theories with this property of Minimal Flavour Violation (MFV, \cite{Buras:2000dm,D'Ambrosio:2002ex}) can have TeV-scale dynamics without being in conflict with observations \cite{Hurth:2008jc}.

Then again, MFV might well be too restrictive an assumption. Concrete models aiming to explain the origin of flavour often bring about a certain amount of non-minimal flavour violation (see {\em e.g.} \cite{Altmannshofer:2009ne,Grinstein:2010ve,Buras:2011ph,Barbieri:2011ci}). Furthermore, the MFV principle does not preclude the presence of flavour-blind CP violating phases \cite{Mercolli:2009ns,Paradisi:2009ey}.
Therefore, testing whether indeed the Yukawa couplings are the only source of flavour breaking and testing whether indeed the CKM phase is the only source of CP breaking are among the major goals of flavour physics in this decade.

The experimental prospects for progress in this direction are excellent, in view of planned dedicated flavour physics experiments like the Super $B$ factories and searches for electric dipole moments (EDMs) or rare $K$ decays;
Moreover, in the near future, important results on $b\to s$ transitions are expected from the LHC. sects.~\ref{sec:1}--\ref{sec:3} will therefore discuss three particularly promising probes of NP at the LHC: the rare leptonic decays $B_{s,d}\to\mu^+\mu^-$, CP violation in $B_s$ mixing, and angular observables in the $B\to K^*\mu^+\mu^-$ decay. Sect.~\ref{sec:4} contains a brief summary of a model leading to interesting signatures in precision flavour experiments: supersymmetry with hierarchical squark masses, effective MFV and flavour-blind phases.

\section{\boldmath $B_{s,d}\to\mu^+\mu^-$}\label{sec:1}

\begin{table}
\begin{center}
\begin{tabular}{lll}
Observable  & SM prediction & experimental bound \\
\hline
BR($B_s\to\mu^+\mu^-$)  & $(3.2 \pm 0.2) \times 10^{-9}$ & $<43 \times 10^{-9}$ \\
BR($B_d\to\mu^+\mu^-$)  & $(0.10 \pm 0.01) \times 10^{-9}$ & $<7.6 \times 10^{-9}$ \\
\end{tabular}
\end{center}
\caption{SM predictions and 95\% C.L. experimental upper bounds on the $B_q\to\mu^+\mu^-$ branching ratios \cite{CDFPubNote9892,Buras:2010pi}.}
\label{tab:bmumu}
\end{table}

\noindent
The decays $B_{s,d}\to\mu^+\mu^-$ are strongly helicity suppressed in the SM and have not been experimentally observed yet; the current upper bounds still lie one/two orders of magnitude above the SM predictions (see table~\ref{tab:bmumu}) but will be improved in the near future by the Tevatron and LHC experiments.

Concerning NP effects in $B_s\to\mu^+\mu^-$, in models where only the SM Wilson coefficient $C_{10}$ receives NP contributions, an order-of-magnitude enhancement of the branching ratio is disfavoured due to constraints from inclusive and exclusive $b\to s\ell^+\ell^-$ transitions on $C_{10}$. In that case, BR$(B_s\to\mu^+\mu^-)\lesssim10^{-8}$ \cite{Bobeth:2010wg}.

Much larger enhancements are possible in principle in models with contributions to the scalar and/or pseudoscalar Wilson coefficients $C_{S,P}$. In two-Higgs-doublet models, a neutral Higgs penguin contributes to the branching fraction with an enhancement factor of $\tan\beta^4$, where $\tan\beta$ is the ratio of Higgs VEVs. In the MSSM, this dependence is even $\tan\beta^6$. Consequently, an upper bound  $\text{BR}(B_s\to\mu^+\mu^-) <  10^{-8}$ would already constrain numerous well-motivated NP scenarios, such as SUSY GUTs with a unification of Yukawa couplings \cite{Altmannshofer:2008vr}.\footnote{On the other hand, even within the MFV MSSM very large $\tan\beta$ would remain a valid possibility if the trilinear couplings are small, such as in gauge mediation scenarios \cite{Altmannshofer:2010zt}.}

An important test of MFV is represented by the measurement of the ratio of the two $B_q\to\mu^+\mu^-$ branching ratios. In MFV (as defined in \cite{D'Ambrosio:2002ex}), it is simply given by
\begin{equation}
\frac{\text{BR}(B_s\to\mu^+\mu^-)}{\text{BR}(B_d\to\mu^+\mu^-)} =
\frac{\tau_{B_s}f_{B_s}^2m_{B_s}}{\tau_{B_d}f_{B_d}^2m_{B_d}}
\left|\frac{V_{ts}}{V_{td}}\right|^2.
\label{eq:sd}
\end{equation}
While a simultaneous enhancement over the SM satisfying (\ref{eq:sd}) would be a strong indication in favour of MFV, a non-SM effect incompatible with (\ref{eq:sd}) would immediately rule out MFV.
\begin{figure}[tbp]
 \centering
 \includegraphics[width=0.75\textwidth]{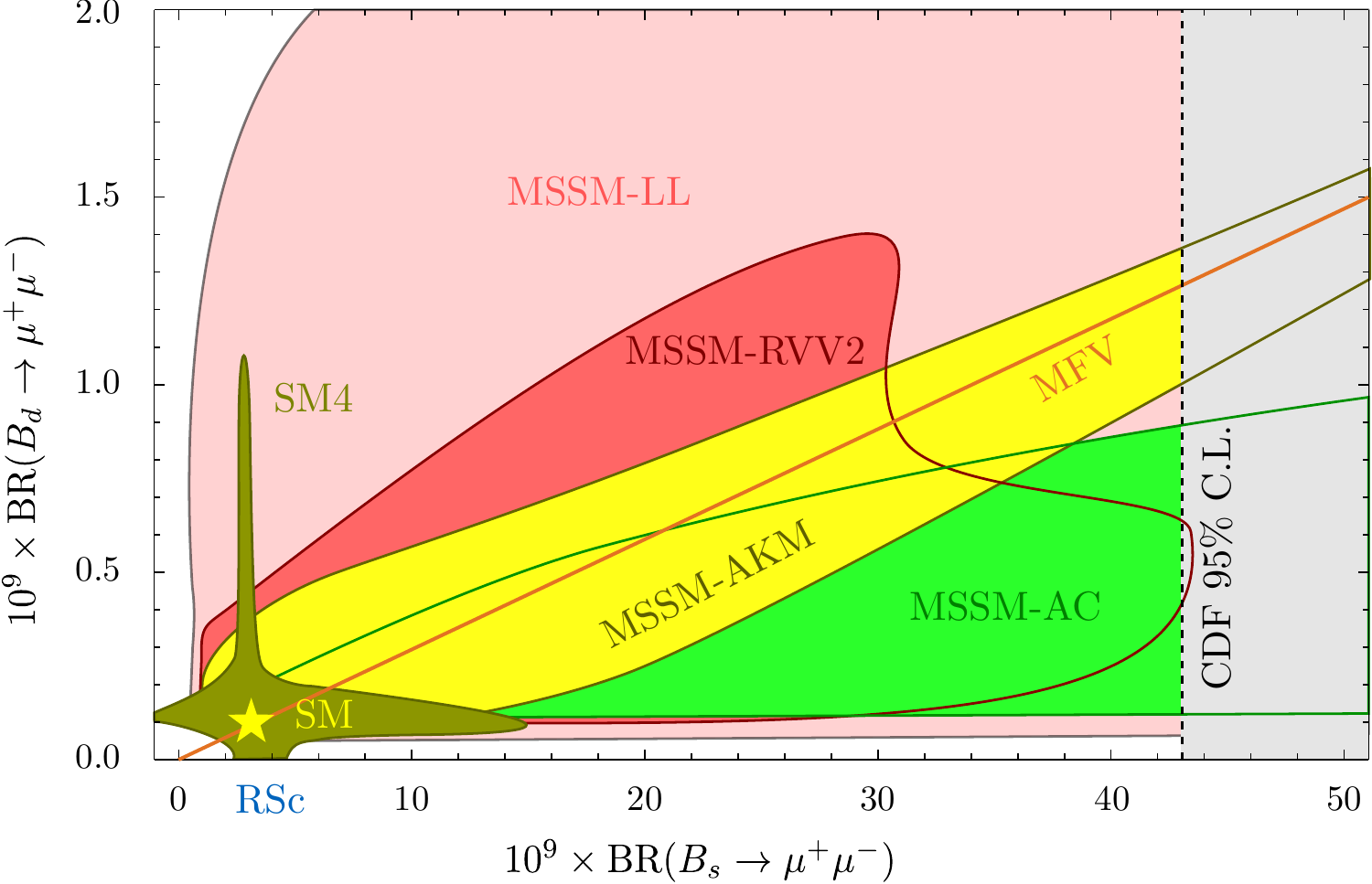}
 \caption{Correlation between the branching ratios of $B_s\to\mu^+\mu^-$ and $B_d\to\mu^+\mu^-$ in MFV, the SM4 and four SUSY flavour models. The gray area is ruled out experimentally. The SM point is marked by a star. Taken from \cite{Straub:2010ih}.}
 \label{fig:bqmumu}
\end{figure}
Figure~\ref{fig:bqmumu} (for details see \cite{Straub:2010ih}) shows possible values of this ratio attained in several non-MFV models: four SUSY flavour models studied in \cite{Altmannshofer:2009ne} and the SM with 4 generations \cite{Buras:2010pi}.
This highlights the power 
of the correlation between $B_s\to\mu^+\mu^-$ and $B_d\to\mu^+\mu^-$
to discriminate between different NP models.

\section{CP violation in \boldmath $B_s$ mixing}\label{sec:2}

In the SM, CP violation in $B_s$ mixing is a small effect since the relevant combination of CKM elements has an accidentally small phase,
\begin{equation}
\phi_s \equiv \text{arg}(M_{12}) = 2\beta_s
\equiv 2\,\text{arg}\!\left(-\frac{V_{ts}^*V_{tb}}{V_{cs}^*V_{cb}}\right)
\approx -0.04 \,.
\label{eq:phis}
\end{equation}
Recently however, two experimental hints for a possibly large non-SM contribution to $\phi_s$ have emerged. One concerns the mixing-induced CP asymmetry $S_{\psi\phi}$ extracted from the time-dependent CP asymmetry in $B_s\to J/\psi\phi$ decays,
\begin{equation}
A^{s}_\text{CP}(\psi\phi,t) 
\equiv \frac{\Gamma(\bar B_s(t) \to \psi\phi) - \Gamma(B_s(t) \to \psi\phi)}{\Gamma(\bar B_s(t) \to \psi\phi) + \Gamma(B_s(t) \to \psi\phi)}
\approx
S_{\psi\phi} \sin(\Delta M_s t)\,,
\end{equation}
where $S_{\psi\phi}=-\sin\phi_s$. The other concerns the charge asymmetry $A_\text{SL}$ in dimuon events at D0, which can be related to the semileptonic CP asymmetries in flavour-specific $B_d$ and $B_s$ decays, $a^{d,s}_\text{SL}$, as \cite{Grossman:2006ce}
\begin{equation}
A_\text{SL}\approx \left(a^d_\text{SL}+a^s_\text{SL}\right)/2 \,,
\label{eq:asl}
\end{equation}
with $O(10\%)$ uncertainties on the coefficients on the right-hand side of (\ref{eq:asl}).

While 2009 results on $S_{\psi\phi}$ showed a discrepancy with the SM somewhere in the ballpark of 3 standard deviations \cite{Punzi:2010nv}, 2010 updates seem to be in agreement with the SM at the $1\sigma$ level \cite{CDFPubNote10206,D0ConfNote6098}, although no combination has been performed yet. The D0 result on $A_\text{SL}$ deviates by $3.2\sigma$ from the SM \cite{Abazov:2010hv}, interestingly pointing in the same direction as the possible effect in $S_{\psi\phi}$\footnote{After this talk was given, preliminary data on $B_s\to J/\psi\phi$ from the LHCb experiment were presented \cite{Uwer} also showing a preference for an effect in the same direction, although still with a small significance.}.

If these hints turn out to be genuine signals of NP in $B_s$ mixing, it would have far-fetching consequences, in particular for theories with MFV. If a theory satisfying MFV does not have any source of CP violation beyond the CKM phase, a sizable $B_s$ mixing phase cannot be generated. Even if the MFV theory has flavour-blind phases, this is nontrivial. While a two-Higgs doublet model with MFV and flavour-blind phases can generate a sizable $\phi_s$ \cite{Kagan:2009bn,Buras:2010mh}, this possibility is precluded in the MSSM by the impact of constraints like $B\to X_s\gamma$ and $B_s\to\mu^+\mu^-$ \cite{Altmannshofer:2009ne}. A confirmation of $\phi_s$ deviating significantly from the value in (\ref{eq:phis}) would thus immediately rule out many well-motivated theories, including the MFV MSSM.

\begin{figure}[tbp]
 \centering
 \includegraphics[width=0.72\textwidth]{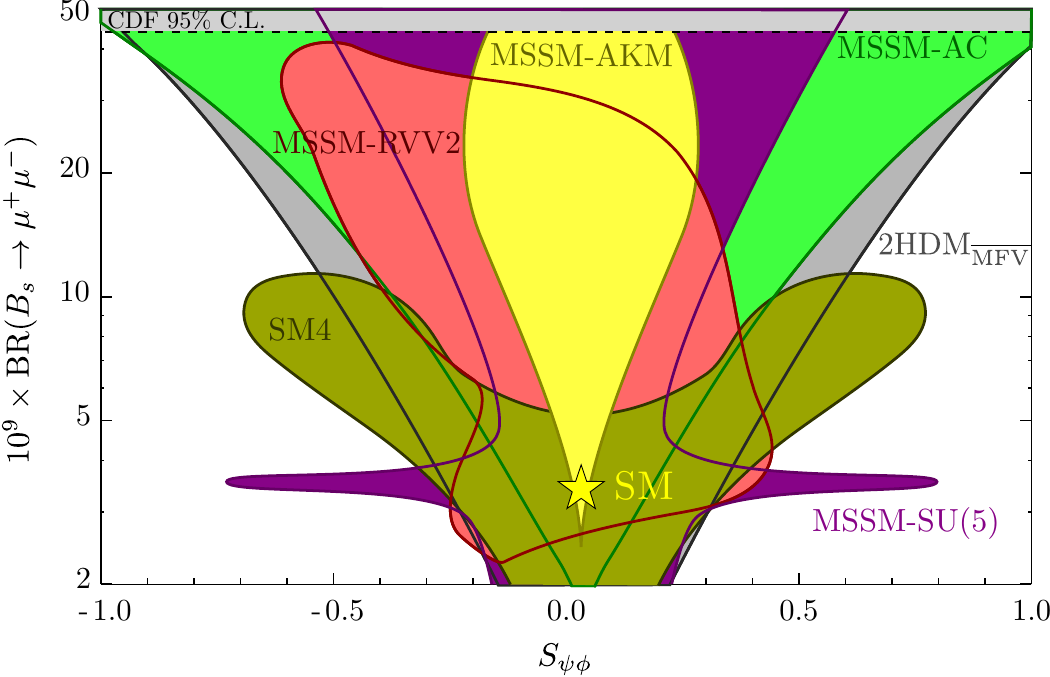}
 \caption{Correlation between the branching ratio of $B_s\to\mu^+\mu^-$ and the mixing-induced CP asymmetry $S_{\psi\phi}$ in the SM4, the two-Higgs doublet model with flavour blind phases and three SUSY flavour models. The SM point is marked by a star.}
 \label{fig:spsiphi}
\end{figure}

Another interesting tool to discriminate between NP models is the correlation between $S_{\psi\phi}$ and the branching ratio of $B_s\to\mu^+\mu^-$. In many models, sizable deviations from the SM prediction for $\phi_s$ are tied to the presence of scalar currents, which can also affect $B_s\to\mu^+\mu^-$. This is the case {\em e.g.} for the two-Higgs doublet model with MFV and  flavour-blind phases (2HDM$_{\overline{\text{MFV}}}$) or for the SUSY flavour model of Agashe and Carone (AC, \cite{Agashe:2003rj}). As shown in fig.~\ref{fig:spsiphi}, sizable $S_{\psi\phi}$ implies a sizable enhancement of BR($B_s\to\mu^+\mu^-$) in these models, while the converse is obviously not true. In the SM4, on the other hand, even the converse statement is true: If $B_s\to\mu^+\mu^-$ is found with a rate significantly enhanced with respect to the SM, this model unambiguously predicts a sizable deviation also in $S_{\psi\phi}$. In yet other models, like in the $SU(5)$ model of \cite{Buras:2010pm}, both observables can be enhanced independently of each other, but a simultaneous enhancement is unlikely.

\section{\boldmath $B\to K^*\mu^+\mu^-$}\label{sec:3}

\noindent
The exclusive decay $\bar B\to \bar K^{*0}(\to K^-\pi^+)\mu^+\mu^-$ gives access to many observables potentially sensitive to NP \cite{Lunghi:2006hc,Bobeth:2008ij,Egede:2008uy,Altmannshofer:2008dz,Bharucha:2010bb,Bobeth:2010wg}; since it is a ``self-tagging'' decay, it also allows a straightforward measurement of CP asymmetries.

The decay poses several theoretical challenges. In addition to calculating the 7 $B\to K^*$ form factors, one has to estimate non-factorizable strong interaction effects. 
At intermediate values of the dilepton invariant mass squared $q^2$, resonant charmonium production leads to a breakdown of quark-hadron duality. The most studied region is the low $q^2$ region, where QCD factorization can be used to calculate non-factorizable corrections \cite{Beneke:2001at,Beneke:2004dp} and light-cone sum rules to calculate the form factors \cite{Ball:2004rg,Altmannshofer:2008dz}.
The high-$q^2$ region above the charmonium resonances has recently attracted increasing attention \cite{Bobeth:2010wg,Beylich:2011aq}. While QCD factorization and LCSR methods are not applicable in this kinematical domain, a local operator product expansion in powers of $1/\sqrt{q^2}$ allows a systematic calculation of the observables \cite{Grinstein:2004vb,Beylich:2011aq}. In \cite{Beylich:2011aq}, it has been argued that, in contrast to the low-$q^2$ region, non-perturbative corrections {\em not} accounted for by the form factors are small (see also \cite{Khodjamirian:2010vf}).

A complete set of observables accessible in the angular distribution of the decay and its CP-conjugate is given by the 9 CP-averaged angular coefficients $S_i$ and the 9 CP asymmetries $A_i$, defined in terms of the angular coefficients $I_i,\bar I_i$ as \cite{Altmannshofer:2008dz}
\begin{equation}
 S_i = \left( I_i + \bar I_i \right) \bigg/ \frac{d(\Gamma+\bar\Gamma)}{dq^2} \,,
\qquad
 A_i = \left( I_i - \bar I_i \right) \bigg/ \frac{d(\Gamma+\bar\Gamma)}{dq^2}\,.
\label{eq:As}
\end{equation}

Not all of the $S_i$ and $A_i$ are both theoretically interesting and experimentally promising. In addition,
at $B$ factories, at the Tevatron and at the early LHC, statistics is quite limited so a full angular analysis is not possible and it might be easier to consider one- or two-dimensional angular distributions depending on a limited set of observables.

In addition to the observables\footnote{See \cite{Straub:2011jy} for a dictionary between different notations for the $B\to K^*\mu^+\mu^-$ observables.} which have already been studied by BaBar \cite{:2008ju}, Belle \cite{:2009zv} and CDF \cite{Aaltonen:2011cn} -- the differential branching ratio, the forward-backward asymmetry and the $K^*$ longitudinal polarization fraction -- two observables sensitive to NP that could be extracted from a one-dimensional angular distribution are the CP-averaged $S_3$ and the CP-asymmetry $A_9$.
While they are both negligibly small in the SM, they could be nonzero in NP models with right-handed currents \cite{Lunghi:2006hc,Altmannshofer:2008dz}.

Two additional observables could be extracted from a two-dimensional angular distribution and might be accessible even during the early LHC running: the CP-averaged observable $S_5$ and the CP-asymmetry $A_7$.
In \cite{Bharucha:2010bb}, it has been shown that even with an integrated luminosity of only $2~\text{fb}^{-1}$, LHCb can measure $S_5$ with a precision that already allows to probe certain NP scenarios. Since $A_7$ is accessible from the same angular distribution, the naive expectation is that the sensitivity should be comparable.
Just as $A_9$, $A_7$ is a T-odd CP asymmetry, meaning that it is not suppressed by small strong phases \cite{Bobeth:2008ij}.
Sizable effects in $A_7$ of up to 20\% are expected in well-motivated NP scenarios like the one discussed in the next section.

\section{CP violation in supersymmetry with effective MFV}\label{sec:4}

\noindent
As mentioned in the introduction, the MFV principle provides a symmetry argument to explain the absence of any evidence for flavour violation beyond the SM, but does not address the absence of signals of CP violation beyond the CKM phase, {\em e.g.} in EDMs.

In supersymmetry, a different way to address the flavour problem is to assume a strong hierarchy between the squarks of the first two versus the ones of the third generation:
the heaviness of the former helps evading the strong bounds from $K$ physics, while the lightness of the latter preserves the SUSY solution to the gauge hierarchy problem. For a generic flavour structure of the soft SUSY breaking terms, this hypothesis is however by far insufficient to ensure the absence of excessive flavour violation and needs to be extended by some amount of flavour alignment (see \cite{Giudice:2008uk} and references therein).

In ref. \cite{Barbieri:2010ar}, the hierarchical sfermion idea was combined with a specific flavour symmetry breaking pattern inspired by MFV. The starting point is a global flavour symmetry in the quark sector of the form
\begin{equation}
G_F = U(1)_{\tilde B_1}\otimes U(1)_{\tilde B_2}\otimes U(1)_{\tilde B_3}\otimes U(3)_{d_R},
\end{equation}
where $\tilde B_i$ acts like baryon number, but only for the right-handed up-type and the left-handed (s)quark fields.
$G_F$ is subsequently broken down to baryon number by a single spurion, the down-type Yukawa coupling $Y_d$. As a consequence, it was shown in \cite{Barbieri:2010ar} that the theory is essentially MFV-like at low energies (therefore the name ``Effective MFV''), with the most significant constraint coming from CP violation in neutral kaon mixing.

Interestingly, hierarchical sfermions also ameliorate the SUSY CP problem compared to MFV since the one-loop contributions to the experimentally accessible EDMs involve the superpartners of first generation fermions.
Therefore, in ref. \cite{Barbieri:2011vn} the Effective MFV (EMFV) framework was analyzed allowing all the CP violating phases not forbidden by the flavour symmetry. With hierarchical squark masses, the usual argument that flavour blind phases need to be tiny to meet the EDM bounds does not apply in EMFV.

The main results of \cite{Barbieri:2011vn} can be summarized as follows,
\begin{itemize}
\item The one-loop contributions to the EDMs are under control if the first generation up-squark and sneutrino masses fulfill the following bounds,
\begin{align}
m_{\tilde{u}} &> {2.7 \mbox{ TeV }} \times (\sin \phi_\mu \tan \beta)^{\frac{1}{2}},
&
m_{\tilde{\nu}} &> {4.0 \mbox{ TeV }} \times (\sin \phi_\mu \tan \beta)^{\frac{1}{2}}.
\end{align}
\item Two-loop contributions from Barr-Zee type diagrams not suppressed by first or second generation sfermion masses give contributions to the electron EDM that are in the ballpark of the experimental bound in the case of a sizable phase in the $\mu$ term, and are usually below it if the stop trilinear coupling (but not $\mu$) is complex.
\item Even if the EDMs are under control, potentially visible effects can arise in CP asymmetries in $B$ physics, induced by one-loop contributions involving third generation sfermions to the magnetic and chromomagnetic $b\to s$ dipole operators.
\end{itemize}

\begin{figure}
\begin{center}
\includegraphics[width=0.48\textwidth]{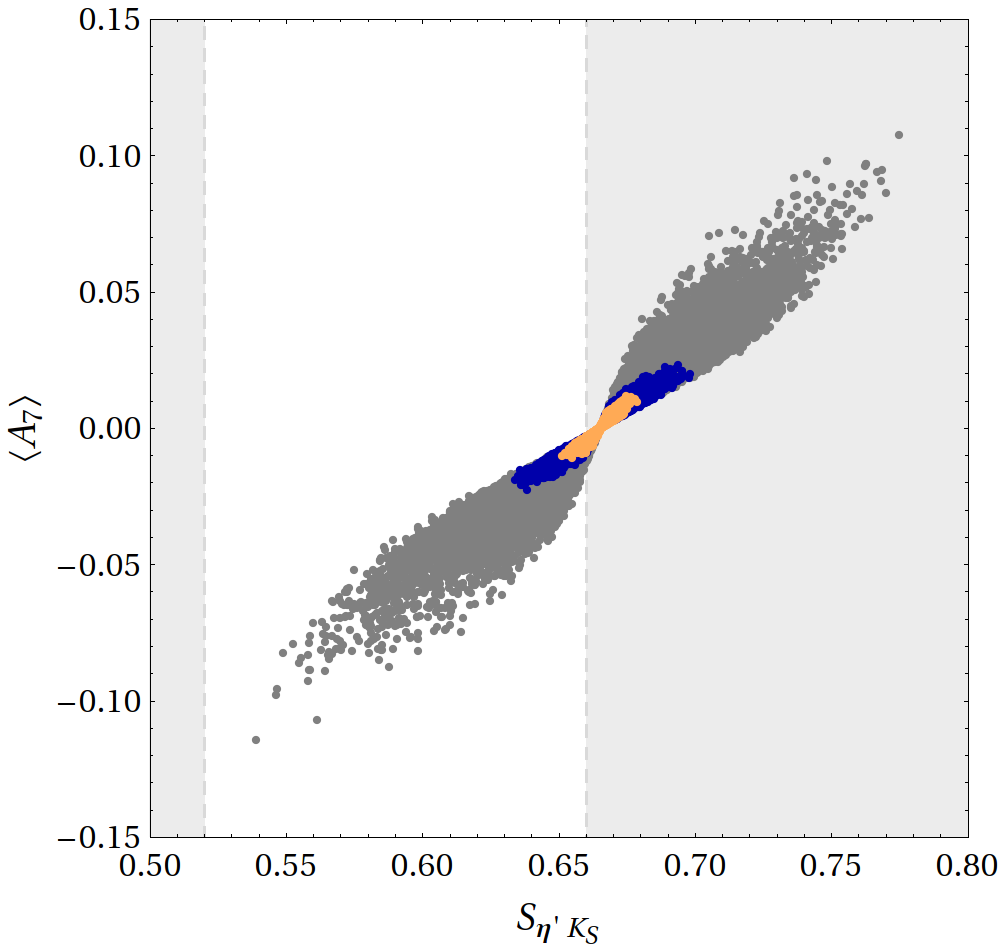}%
\hspace{0.019\textwidth}%
\includegraphics[width=0.465\textwidth]{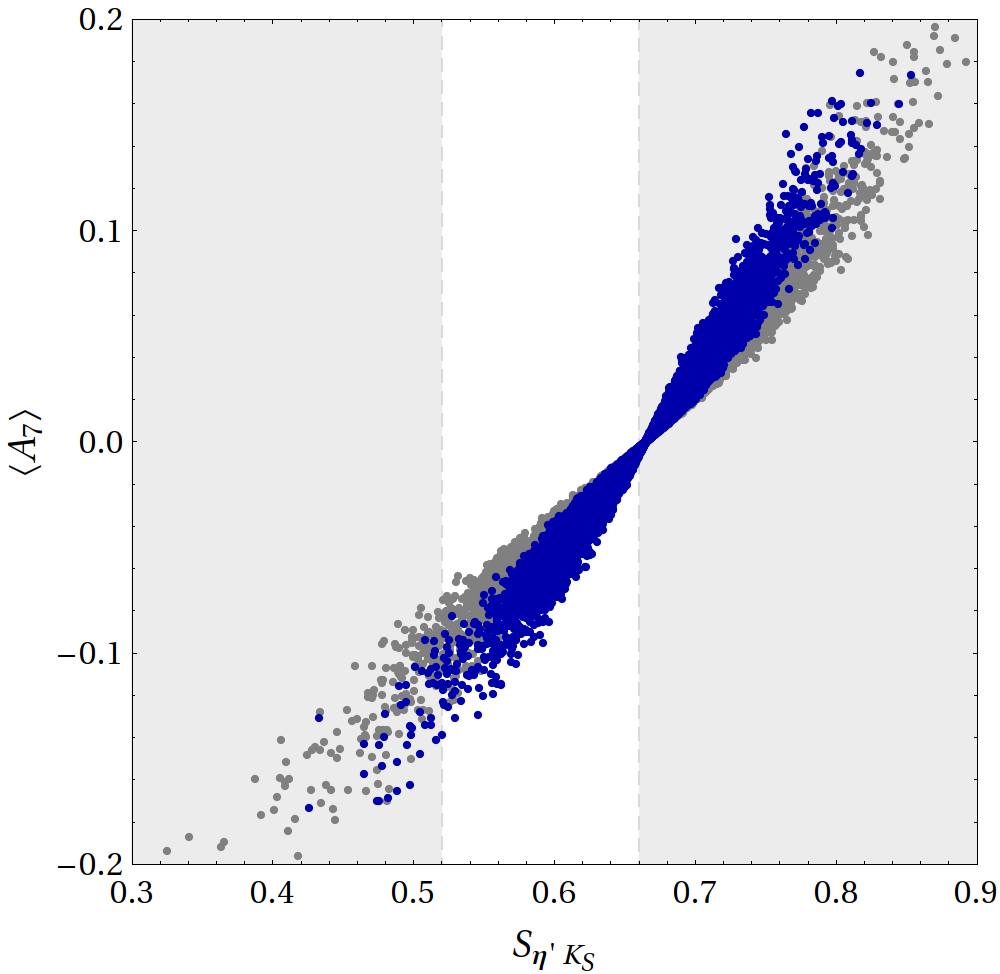}
\end{center}
\caption{Correlation between the mixing-induced CP asymmetry in  $B\to\eta' K_S$ and the angular CP asymmetry $\langle A_7 \rangle$ in $B\to K^*\mu^+\mu^-$ in two scenarios with a complex $\mu$ term (left) or complex $A_t$ term (right).
The gray points are allowed by all constraints except $d_e$, while the blue points are compatible with all constraints. The orange points in the left-hand plot have $|\sin\phi_\mu|<0.2$.}
\label{fig:SvsA}
\end{figure}

As an example for the possible effects in $B$ physics, fig.~\ref{fig:SvsA} shows the correlation between the mixing-induced CP asymmetry in  $B\to\eta' K_S$ and the angular CP asymmetry $A_7$ (integrated in the low $q^2$ range) in $B\to K^*\mu^+\mu^-$ discussed in sect.~\ref{sec:3} for two scenarios: first, assuming an arbitrary phase of the $\mu$ term, second, assuming an arbitrary phase for the stop trilinear coupling. In particular in the second scenario, sizable effects in the CP asymmetries can be generated without violating the EDM bounds\footnote{The signals in flavour 
physics arising in this second scenario are very similar to the effects in the MFV MSSM 
with a complex $A_t$ term and a real $\mu$ term \cite{Altmannshofer:2008hc,Altmannshofer:2009ne}. Of course, the two setups are easily distinguishable on the basis of their different spectrum.}.

This study provides an example of a model with interesting signatures in flavour physics observables and also highlights the importance of dedicated flavour physics experiments, like EDM searches or Super $B$ factories, as complements to the LHC. The results are applicable to all MSSM scenarios with hierarchical sfermions (see {\em e.g.} \cite{Barbieri:2011ci}), with possible additional effects in the presence of non-minimal flavour violation.

\section{Conclusions}

\noindent
The LHC era is in full swing, also in flavour physics. Imminent results on decays like $B_s\to\mu^+\mu^-$ or $B\to K^*\mu^+\mu^-$ and on CP violation in $B_s$ mixing discussed in this talk will help shed light on the question whether there is new physics at the TeV scale, and subsequently whether the Yukawa couplings are the only source of breaking of the flavour and CP symmetries, as in the SM.

Supersymmetry with hierarchical squark masses, combined with Effective MFV \cite{Barbieri:2010ar,Barbieri:2011vn}, can solve the SUSY flavour {\em and} CP problems and leads to potentially visible effects in CP asymmetries in $B$ physics, without violating EDM bounds.

\acknowledgments
\noindent
I thank the organizers for their invitation to this wonderful conference and Wolfgang Altmannshofer for useful comments on the manuscript.

This work was supported by the EU ITN ``Unification in the LHC Era'', contract PITN-GA-2009-237920 (UNILHC).

\end{document}